\documentclass[twocolumn,preprintnumbers,amsmath,amssymb,prl,floatfix,showpacs]{revtex4} 
\usepackage{epsfig}
\usepackage{physics}
\begin{document}
\title{Two dimensional electron gas in a non-Euclidean space}

 \author{Raimundo N. Costa Filho} 
\email{rai@fisica.ufc.br} \affiliation{Departamento de F\'isica,
  Universidade Federal do Cear\'a, Caixa Postal 6030, Campus do Pici,
  60455-760 Fortaleza, Cear\'a, Brazil}
  
  \author{S. F. S. Oliveira}
\email{stanleyfrota@gmail.com}
\affiliation{Departamento de F\'isica,
  Universidade Federal do Cear\'a, Caixa Postal 6030, Campus do Pici,
  60455-760 Fortaleza, Cear\'a, Brazil}

\author{V. Aguiar}
\email{vanderley.junior.ufc@gmail.com} \affiliation{Departamento de F\'isica,
  Universidade Federal do Cear\'a, Caixa Postal 6030, Campus do Pici,
  60455-760 Fortaleza, Cear\'a, Brazil}

\date{\today}

\begin{abstract}

A charged particle in the presence of a magnetic field is studied in the position dependent operator formalism. Instead of a quantum harmonic oscillator,  the solution of the resulting Schr\"odinger-like equation is the one for the Morse oscillator. The anharmonicity that shows up naturally from the theory is analogous to the corrections introduced by relativistic ones.The degeneracy of the spin up and down levels is lifted due to the non-Euclidean space. It is shown that the  Fermi energy  and the total ground state energy of the 2DEG is also modified.

\end{abstract}
\maketitle
 
In a two dimensional electron gas with a perpendicular applied magnetic field the electrons are trapped in orbits and move with a  frequency that depends  on the applied magnetic field. The electron motion is quantized with energies given by $E_n=\hbar\omega_c(n+1/2)$, where $\omega_c$ is the cyclotron frequency of the electron. The levels are equally spaced and separated by $\Delta E=\hbar\omega_c$. These are the well known Landau Levels (LLs). The density of states of these LLs are delta functions. However, the experiments show that the  in real systems various scattering processes give rise to an energy spread in the Landau levels, resulting from the finite lifetime of electrons in a given state. Even when such effect are not considered, the LLs are split due to the spin. This spin splitting is introduced through the Zeeman term in the Hamiltonian where each Landau level splits into a pair, one for spin up electrons and the other for spin down electrons.

Despite its obvious applicability in semiconductors theory and the source to the discover of the quantum Hall effect\cite{Kvk}, the above described system has been used to test theories  outside the realm of condensed matter.  More precisely, quantum gravity and black holes\cite{Das}.  Besides using condensed matter systems as a lab test, these theories has is common the fact they used a more general Heisenberg uncertainty.
 
The Heisenberg uncertainty principle is one of the fundamental relation of the Quantum Mechanics. This relation imposes a lower limit for the product of the uncertainties in position and momentum spaces; the determination of the exact position of a particle requires an infinite momentum uncertainty. Despite its importance in the conventional Quantum Mechanics, the Heisenberg relation should be replaced by a more general relation in the limit of  distances of the order of Planck length $(l_P=\sqrt(G\hbar/c^3 )\approx10^{-35} m)$ where gravitational effects play an important role \cite{Hossenfelder}. 

A possible way to include the effects of gravity in quantum mechanics  is to  modify the canonical commutation relation between position and momentum operators considering that there exists a minimal length. That leads to the Generalized uncertainty principle (GUP) $\Delta  x\Delta p\ge\hbar[1+\alpha^al_p^2(\Delta p)^2/\hbar]$. For instance, S. Benczik et al \cite{Benczik} studied the effects of GUP in the Hydrogen atom and obtained the energy spectrum of the Coulomb potential both numerically and perturbatively for arbitrary values of $\alpha$. Pedram \cite{Pedram2} showed that the presence of the maximal momentum results in upper bounds on the energy spectrum of the free particle and the particle in box. In 2018, Bosso \cite{Bosso} derived the necessary form of the Legendre transformation to connect Hamiltonian and Lagrangian functions with a minimal length. He analyzed the harmonic oscillator and Kepler's problem, and how these potentials are transformed from one formalism to the other. Other interesting works about GUP can be found in references\cite{Lubo,Scardigli,Rossi,Zhao,Castro, Moniruzzaman}.

In addition to a minimum length, a minimum momentum has been included in the generalized uncertainty principle to turn it more symmetric.  It turns out that this term is called the extended uncertainty principle $\Delta x\Delta P_x\ge  \hbar[1+\beta^2(\Delta x)^2/l^2]$ where $\beta$ is a constant and $l$ is a length scale and it has been proposed to investigate deSitter black holes thermodynamics \cite{Bolen,Mignemi,Kempf,Ashgari}.

Recently, Costa Filho et al. \cite{Costa} were able to derive  the extended uncertainty principle (EUP) from first principles. They consider a distance between two points given by $ds^2=\sum_{\mu}g_{\mu\mu}d\mu d\mu$, where $\mu=x,y,z$. $g_{\mu\mu}$ is an element of a diagonal metric of the non-Euclidean space considered. In this space, a particle goes from point $x$ to $x+g_{xx}^{-1/2}dx$ through the translation operator $T_g (dx)\ket{x}=\ket{x+ g_{xx}^{-1/2}dx}$, where $g_{xx}^{-1/2}$ is a function of position can be Taylor expanded as $g_{xx}^{-1/2}=\sum_{n=0}^{\infty}a_nx^n$ \cite{Costa1,Costa2}. 

In this context of the position dependent translation operator  (PDTO) formalism, the above-mentioned problem of an electron moving  in the presence of a magnetic field  is studied using the Hamiltonian 
\begin{equation}
\label{Hamiltonian}
H = \frac{1}{2m_e}\left({\bf \mathcal{P}} + e{\bf A}\right) \cdot \left({\bf \mathcal{P}}+e{\bf A}\right),
\end{equation}
with $\mathcal{P}_\nu = -i\hbar g_{\nu\nu}^{-1/2}\frac{\partial}{\partial \nu}$ \cite{Costa1}, where the $g_{ij}$ is the metric  under consideration.

It is important to show that Eq. (\ref{Hamiltonian}) is invariant under a transformation of the vector potential
\begin{equation}
    \bf{A'} = \bf{A} - \bf{\nabla} \Lambda, 
\end{equation}
where $\Lambda = \Lambda\left(x, y\right)$. Taking the equation below
\begin{equation}
    \left(-i\hbar g_{\nu\nu}^{-1/2}\frac{\partial}{\partial \nu} +eA'_{\nu} \right) U(x, y, t)\psi(x, y, t)
\end{equation}
where $\psi(x, y, t)$ is the eigenfunction of the Hamiltonian (\ref{Hamiltonian}) and $U(x, y, t)$ is the operator that transforms $\psi$ into $\psi'$, {\it i.e.}, $\psi' = U\psi$. One can use  $U = u(t)e^{f(x)}e^{h(y)}$ to show that
\begin{equation}
   \left(-i\hbar g_{\nu\nu}^{-1/2}\frac{\partial}{\partial \nu} +eA'_{\nu} \right) U\psi = U\left(-i\hbar g_{\nu\nu}^{-1/2}\frac{\partial}{\partial \nu} +eA_{\nu} \right)\psi,
\end{equation}
for
\begin{equation}
    \frac{df}{dx}=\frac{e}{i\hbar}\frac{1}{g_{xx}^{-1/2}}\frac{\partial \Lambda}{\partial x},
\end{equation}
and
\begin{equation}
    \frac{dh}{dy}=\frac{e}{i\hbar}\frac{1}{g_{yy}^{-1/2}}\frac{\partial \Lambda}{\partial y},
\end{equation}
thus, making an appropriate change from $\psi$ to a newly defined $\psi'$ we can restore the equation to its original form, and thus show the gauge invariance of the particle's equation of motion.

Consider now the Landau gauge ${\bf {A}} = (0, Bx, 0)$, Eq. (2) becomes
\begin{align}
    H = &-\frac{\hbar^2}{2m_e}g_{xx}^{-1/2}\frac{\partial}{\partial x}\left(g_{xx}^{-1/2}\frac{\partial}{\partial x}\right) + \frac{m_e\omega_c ^2 x^2}{2} \nonumber \\
    &-\frac{\hbar^2}{2m_e}g_{yy}^{-1/2}\frac{\partial}{\partial y}\left(g_{yy}^{-1/2}\frac{\partial}{\partial y}\right) - i\hbar\omega_c x g_{yy}^{-1/2}\frac{\partial}{\partial y},
\end{align}
where $\omega_c = eB/m_e$ is the cyclotron frequency.

The eigenvalue equation for Hamiltonian (7), $H\Psi(x,y)=E\Psi(x,y)$, writing $\Psi(x,y)=e^{ik_y \int{g_{yy}^{1/2}dy}} \psi(x)$ reads
\begin{align}
&-\frac{\hbar^2}{2m_e}g_{xx}^{-1/2}\frac{d}{dx}\left(g_{xx}^{-1/2}\frac{d}{dx}\right)\psi(x) + \nonumber \\
&\frac{m_e\omega_c^2}{2}\left(x+\frac{\hbar k_y}{m_e\omega_c}\right)^2 \psi(x) = E\psi(x).
\end{align}
The solution can be written as
\begin{equation}
    \Psi(x,y) = e^{ik_y \int{g_{yy}^{1/2}dy}}\phi(x-x_0),
\end{equation}
where $x_0 = -\hbar k_y/m_e\omega_c=l_0^2k_y$,  and $l_0=\sqrt{\hbar/eB}$ is a length scale called magnetic length.

For a metric expanded to the first order in $x$, $g_{xx}^{-1/2} = \left[1+\gamma\left(x-x_0\right)\right]$. Here we define $\gamma=\sqrt{g^*}/l_0$ where $g^*$ is a dimensionless constant, and the appropriated length scale for the problem is the magnetic length. 

\begin{figure}
\centering
\includegraphics[scale=0.4]{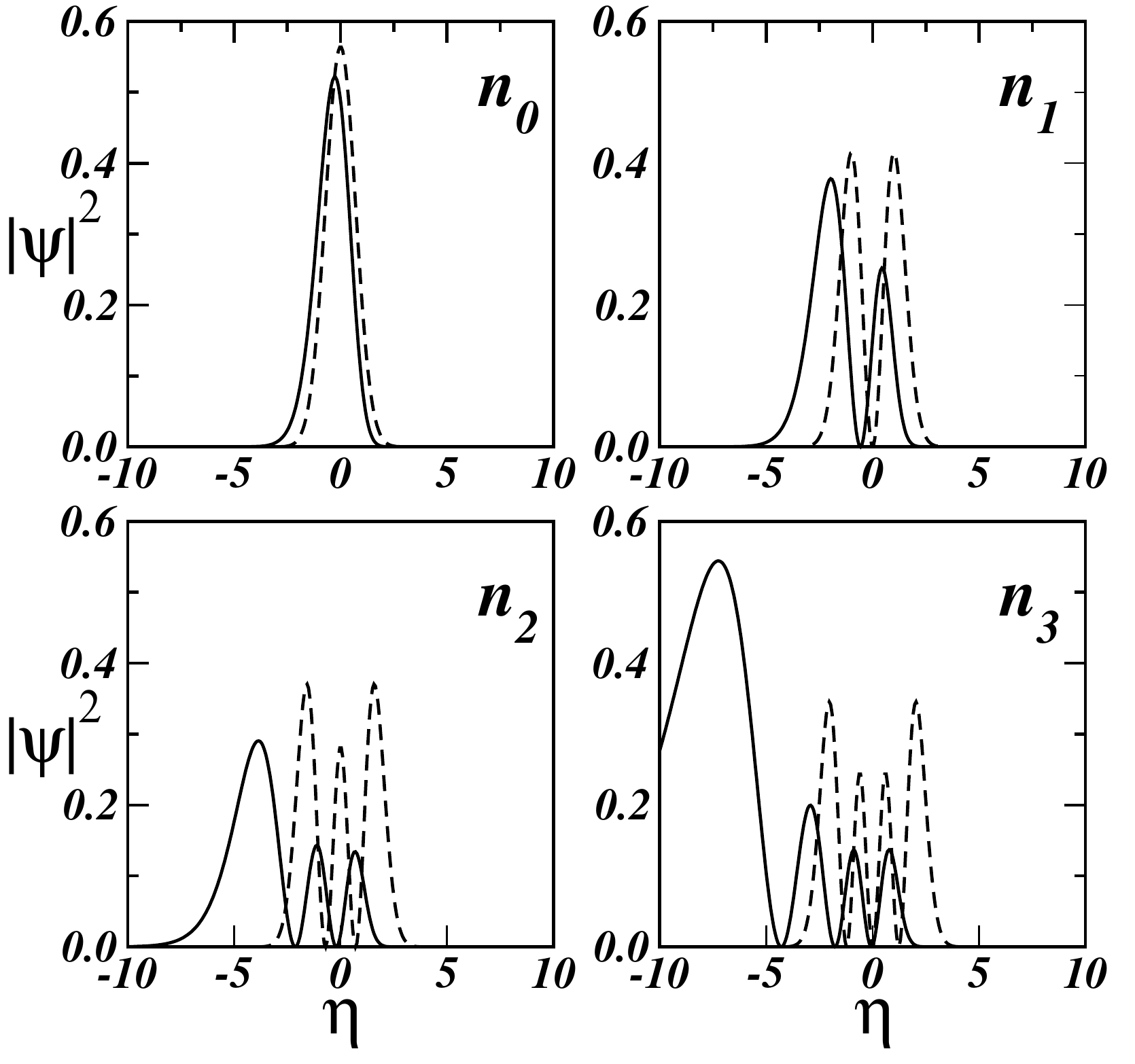}
\caption{ The wavefunction amplitude for the first four levels. The solid line is for the parameter $g^* =0.25$, the dashed line is for $g^* =0$. The magnetic field used was $B=1T$}\label{fig1}
\end{figure}

\begin{figure}
\centering
\includegraphics[scale=0.3]{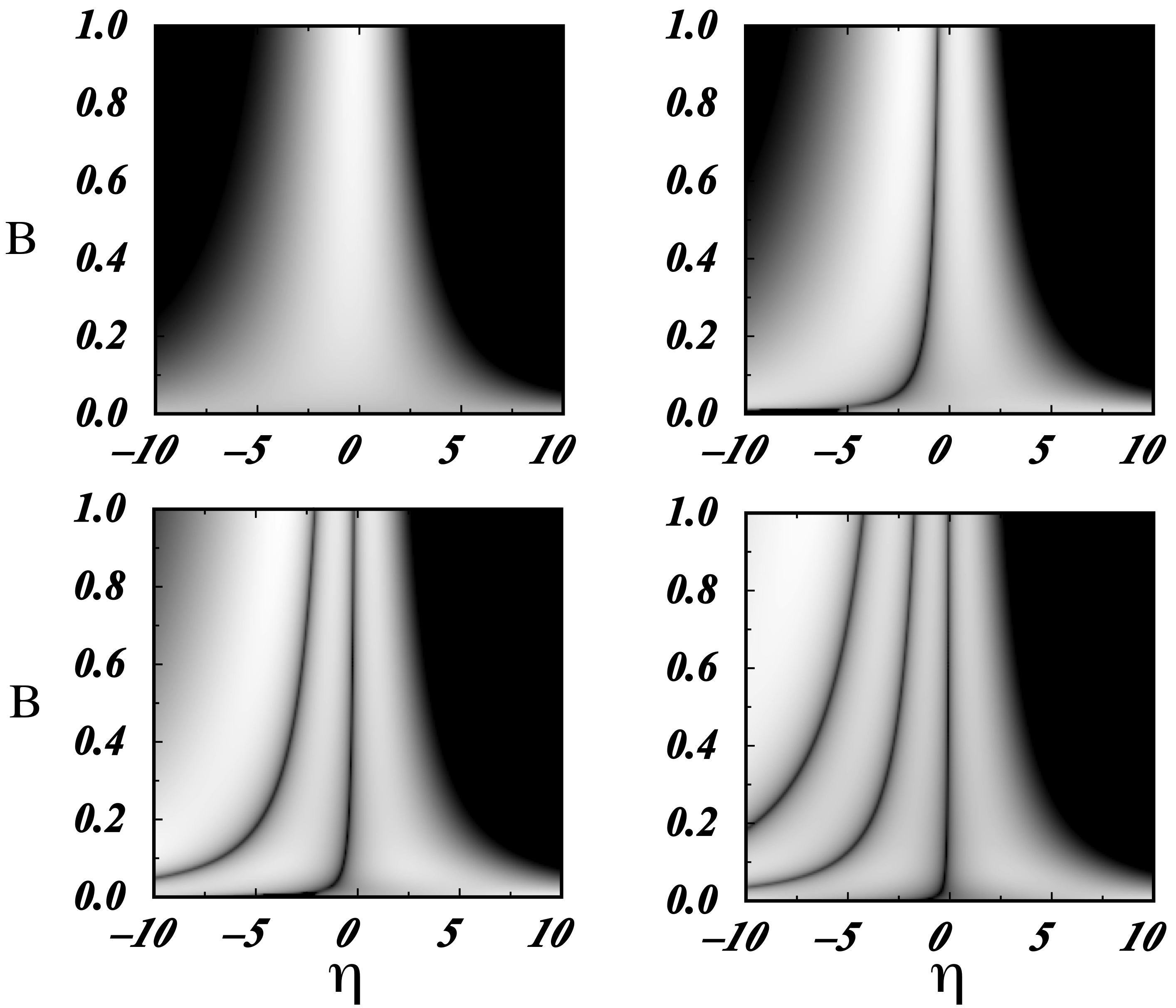}
\caption{ The dependence of the wave function on the magnetic field. The amplitude increases from black to white.}\label{fig2}
\end{figure}

The eigenfunction is given by
\begin{align}
    \Psi(\eta,y) = &A_n e^{ik_y \int{g_{yy}^{1/2}dy}} \left(\frac{2}{g^*}e^{\sqrt{g^*}\eta/l_0}\right)^{s}\times \nonumber\\
    &\exp\left[-\frac{1}{2}\left(\frac{2}{g^*}e^{\sqrt{g^*}\eta/l_0}\right)\right] L_{n}^{2s}\left(\frac{2}{g^*}e^{\sqrt{g^*}\eta/l_0}\right),
\end{align}
where $\eta =(l_0/\sqrt{g^*})\ln\left[1+\sqrt{g^*}\left(x-x_0\right)/l_0\right]$, $s=1/g^*- \left(n+1/2\right)$, $L_{n}^{2s}$ is the generalized
Laguerre polynomial, and $A_n$ is the normalization constant.  The wave function  amplitude depends on $g^*$ and $B$. For two different values of $g^*$ and a fixed value of the magnetic field, Figure \ref{fig1} shows that dependence. For $g^* =0$, the system's wave function is exactly the one for the  quantum harmonic oscillator, while considering $g^*=0.25$ the wave function loses its symmetry around $\eta=0$. In Figure \ref{fig2} a contour plot of the wave function amplitude is shown for a fixed value of $g^*$ with a varying magnetic field. As the magnetic field increases the particle gets more confined.


The eigenvalues of the above functions are the energies
\begin{equation}\label{energy}
    E_{n} = \hbar \omega_c \left(n + \frac{1}{2}\right)\left[1 -\frac{g^*}{2}\left(n + \frac{1}{2}\right)\right],
\end{equation}
 As one can see, the levels are not equally spaced and the difference between them decreases as $n$ increases. 
\begin{equation}
    \Delta E_{n+1,n} = \hbar \omega_c\left[1 -g^*(n+1)\right].
\end{equation}
The number of electronic states of the 2DEG without the magnetic field has to be equal to the number of states when the field is turned on. Therefore, one can show that the number of electrons per unit area in a level $n$ is given by

 \begin{equation}
\mathcal{N}_B=\frac{B}{\phi_0}[1-g^*(n+1)],
\end{equation}
where $\phi_0$ is a flux quantum for a pair of electrons. The filling factor is given by
\begin{equation}
\nu=\frac{\mathcal{N}_B}{\mathcal{N}_{0}},
\end{equation}
here $ \mathcal{N}_{0}$ is the number of electrons per unit area without the magnetic field. When the filling factor is equal to an integer $i$ it means that all the levels up to that integer $i$ will be filled. That happens everytime $B=i [1-g^*(n+1)]B_0$, where $B_0=\mathcal{N}_{0}\phi_0$.

\begin{figure}
\centering
\includegraphics[scale=0.4]{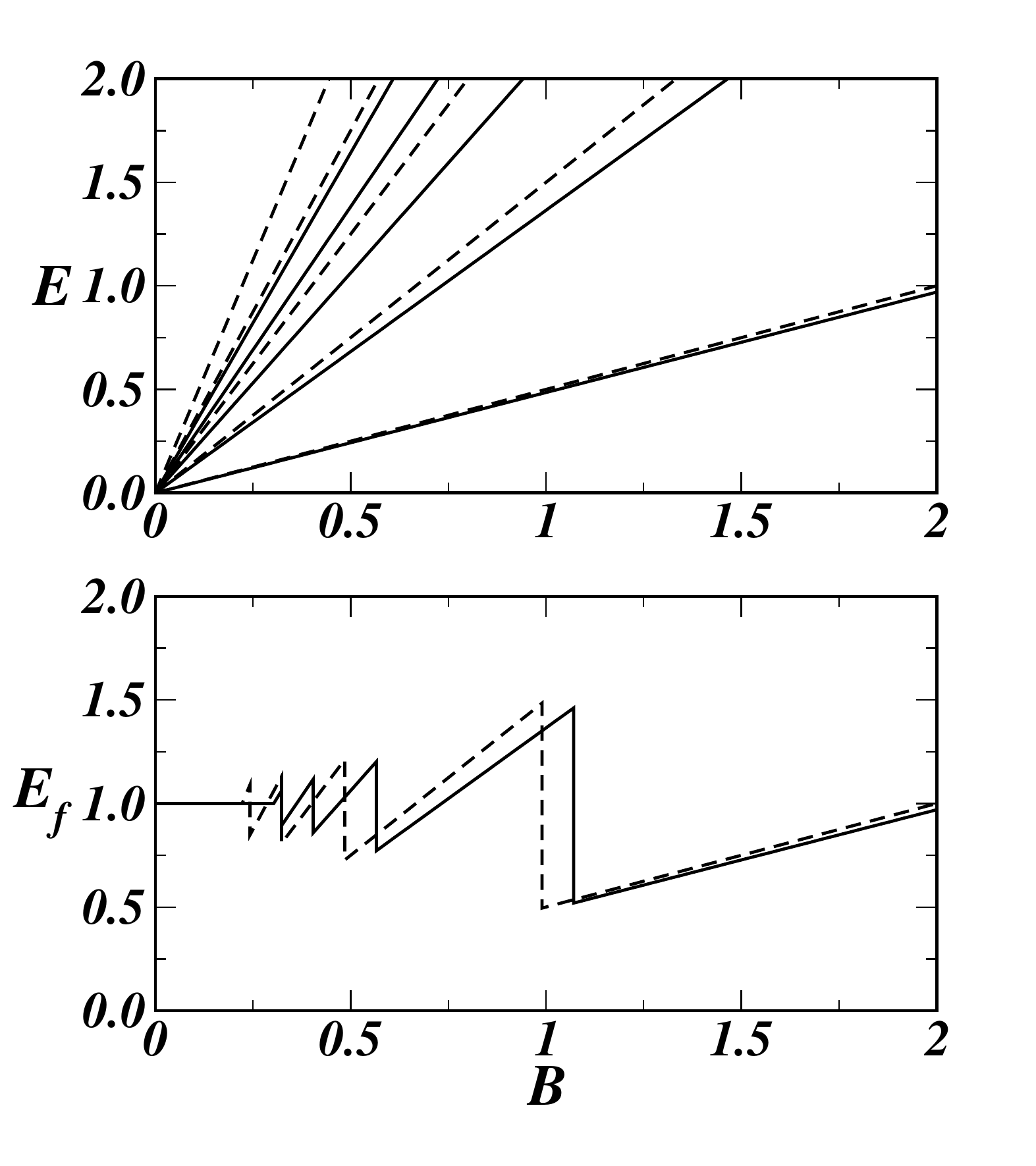}
\caption{ The dependence of the energy levels (upper panel) and Fermi energy (lower panel) on the magnetic field. The solid lines shows the energies for $g^*=0.12$, and the dashed lines for $g^*=0$.}\label{fig3}
\end{figure}

It is clear that the energy levels are not equally spaced for $g^*>0$. That is shown in the upper panel of Fig. \ref{fig3}, where the energy levels are plotted against the magnetic field. The dashed lines are the usual LL's. In the lower panel of the figure one can see the behavior of the 2DEG Fermi energy against the magnetic field.  For strong magnetic fields only the lowest energy level is populated. As the field decreases the other levels are getting populated. The main effect of the modified space is the shift of Fermi energy jumps. 

\begin{figure}
\centering
\includegraphics[scale=0.4]{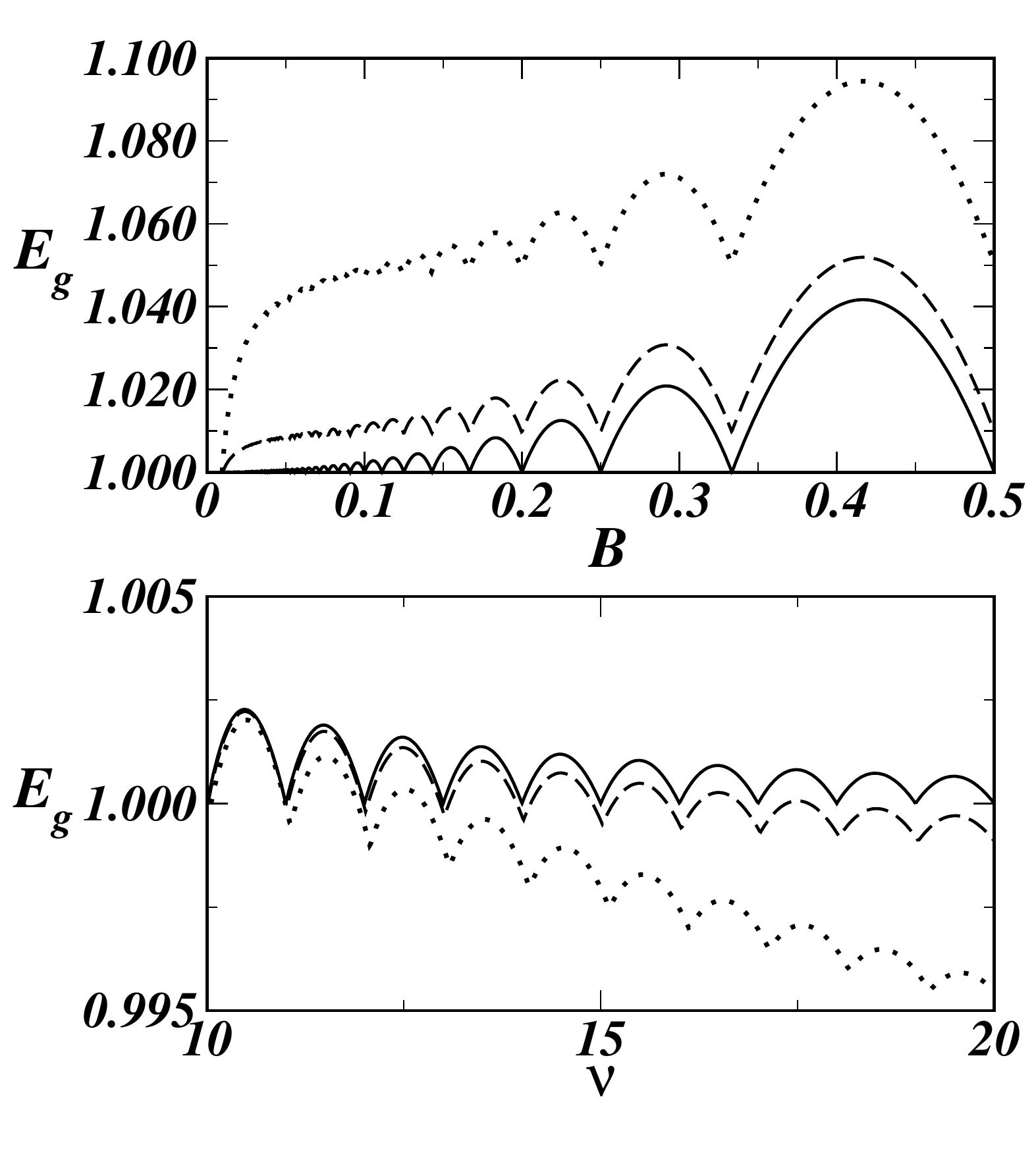}
\caption{ The dependence of the 2DEG ground state energy on the magnetic field  $B$ (upper panel) and the filling $\nu$ (lower panel). The solid lines shows the energies for $g^*=0$, while the dashed and dotted lines are for $g^*=10^{-4}$ and $g^*=5\times10^{-4}$, respectively.}\label{fig4}
\end{figure}

The effect of modified space is even more pronounced for the ground state energy of the system. Figure \ref{fig4} shows the oscillations of the ground state energy against the magnetic field in the upper panel, and against the filling factor in the lower panel. The solid lines  in both panels are for $g^* =0$. The amplitude oscillations increases as $B^2$ and its period gets
smaller as $B$ decreases.  The dashed and dotted lines are for $g^*=10^{-4}$ and $g^*=5\times10^{-4}$, respectively. For small values of the magnetic field, and tiny values of $g^* $, the oscillations are meaningless and the energy grows till it starts oscillating for larger values of $B$. In the lower panel the oscillations decreases with the filling factor. For integer values of $\nu$, the ground state gets a minimum of energy.   From the Figure \ref{fig4} one can see that this minimum decreases with the filling factor fo $g^*>0$.

 The above results change the electronic and magnetic properties of the two dimensional electron gas. They are not related to a quantum harmonic oscillator as in the case of a flat space. Instead, they are the eigenfunctions and eigenvalues of the Morse potential $V=D(1-e^{-\alpha x})^2$. This is an anharmonic potential used to describe the dissociation of diatomic molecules.  In the PDTO formalism the harmonic oscillator has in fact the energies and wave function of the Morse Potential \cite{Costa-Morse}. One important characteristic of such system is that, as one can see from the energy expression, there is a finite number of quantized states that depends on the condition  $n\le1/g^*-1$. 
 
 It is important to remember that $g^*$ is a dimensionless parameter related to the metric used and is related to the anharmonicity of the oscillator.  The anharmonicity that comes from the modified space is analogous to the anharmonicity provoked by relativistic effects. Relativistic corrections introduces anharmonicity through a relativistic mass increase $m=(1+K/mc^2)m_e$, where $K$ is the electron's kinetic energy. In our formalism the effective mass can be written as
 
 \begin{equation}
 m=\frac{m_e}{1-\frac{g^*}{2}\left(n + \frac{1}{2}\right)}.
 \end{equation}
 And the shift in the energy levels from the relativistic corrections $\Delta E/\hbar=\omega_c-\delta(n+1)$ \cite{Gabrielse,Gabrielse1}(not considering the spin contribution) is just the result obtained here with $g^*=\delta/\omega_c$.

 In summary we have shown that the position dependent translation operator formalism (PDTO) that leads to an extended uncertainty principle (EUP) is able to give corrections to the electron mass and consequently to the energy levels that are analogous to the relativistic corrections.  The corrections come directly from the theory without any ad-hoc inclusion of extra terms in the Hamiltonian. We have used the metric $g_{xx}^{-1/2}=1+\sqrt{g^*}x/l_0$, with a characteristic length defined as the magnetic length of the system. For $g^*>0$, the system gets anharmonic described by the Morse potential. The energy levels are not equally spaced, leading to the lift of the spin-up spin-down degeneracy. It also changes the fermi energy, and  the total ground state energy oscillations. That will be observed in the quantum hall effect, and in magnetic oscillations like de Haas-van Alphen and Shubnikov-de Haas oscillations.

The authors are grateful to the National Counsel of Scientific and Technological Development (CNPq) and to the National Council for the Improvement of Higher Education (CAPES) of Brazil for financial support. RNCF is supported by CNPq grant number 312384/2018-1


\begin{thebibliography}{}
\bibitem{Kvk}K. v. Klitzing, G. Dorda, M. Pepper, Phys. Rev. Lett. {\bf45}, 494 (1980).
\bibitem{Das}S. Das and E. C. Vagenas, Phys. Rev. Lett. {\bf101}, 221301 (2008).
\bibitem{Hossenfelder} S. Hossenfelder, Living Rev. Relativity {\bf 16}, 2 (2013).
\bibitem{Benczik} S. Benczik, L. N. Chang, D. Minic and T. Takeuchi, Phys. Rev. A {\bf 72}, 012104 (2005).
\bibitem{Pedram2} P. Pedram, Phys. Lett. B {\bf 718}, 638 (2012).
\bibitem{Bosso} P. Bosso, Phys. Rev. D {\bf 97}, 126010 (2018).
\bibitem{Lubo} M. Lubo, Phys. Rev. D {\bf 61}, 124009 (2000).
\bibitem{Scardigli} F. Scardigli and R. Casadio, Eur. Phys. J. C  {\bf 75}, 425 (2015).
\bibitem{Rossi} M. A. C. Rossi, T. Giani and M. G. A. Paris, Phys. Rev. D {\bf 94}, 024014 (2016).
\bibitem{Zhao}  Q. Zhao, M. Faizal and Z. Zaz, Phys. Lett. B {\bf 770}, 564 (2017).  
\bibitem{Castro} L. B. Castro and A. E. Obispo, J. Phys. A: Math. Theor. {\bf 50}, 285202 (2017). 
\bibitem{Moniruzzaman}  M. Moniruzzaman and S. B. Faruque, J. Sci. Res. {\bf 10} (2), 99 (2018). 

\bibitem{Bolen}B. Bolen, M. Cavagli\'a, Gen. Relativity Gravitation {\bf 37}, 1255 (2005).
\bibitem{Mignemi}S. Mignemi, Modern Phys. Lett. A {\bf 25}, 1697 (2010).
\bibitem{Kempf}A. Kempf, J. Math. Phys. {\bf 35}, 4483 (1994).
\bibitem{Ashgari}M. Ashgari, P. Pedram, K. Nozari, Phys. Lett. B {\bf 725}, 451 (2013).
\bibitem{Costa}R. N. Costa Filho, M. P. Almeida, G. A. Farias and J. S. Andrade Jr., Phys. Rev. A {\bf 84}, 050102(R) 2011.
\bibitem{Costa1}R. N. Costa Filho, J. P. M. Braga, J. H. S. Lira and J. S. Andrade Jr., Phys. Lett. B {\bf 755}, 367 (2016).
\bibitem{Costa2}J. P. M. Braga and R. N. Costa Filho, Int. J. Mod. Phys. C {\bf 27}, 1650047 (2016).

\bibitem{Costa-Morse}R. N. Costa Filho, G. Alencar, B. S. Skagerstam and J. S. Andrade, EPL, {\bf 101}, 10009 (2013).
\bibitem{Gabrielse} B. Odom, D. Hanneke, B. D'Urso and G. Gabrielse,
Phys. Rev. Lett. 97, 030801 (2006).
\bibitem{Gabrielse1}L.S. Brown and G. Gabrielse, 
Rev. Mod. Phys. 58, 233-311 (1986).
\end{thebibliography}
\end{document}